# Describing Papers and Reviewers' Competences by Taxonomy of Keywords

Yordan Kalmukov

Department of Computer Systems and Technologies,
University of Ruse,
8 Studentska Str., 7017 Ruse, Bulgaria,
JKalmukov@gmail.com

**Abstract**. This article focuses on the importance of the precise calculation of similarity factors between papers and reviewers for performing a fair and accurate automatic assignment of reviewers to papers. It suggests that papers and reviewers' competences should be described by taxonomy of keywords so that the implied hierarchical structure allows similarity measures to take into account not only the number of exactly matching keywords, but in case of non-matching ones to calculate how semantically close they are. The paper also suggests a similarity measure derived from the well-known and widely-used Dice's coefficient, but adapted in a way it could be also applied between sets whose elements are semantically related to each other (as concepts in taxonomy are). It allows a non-zero similarity factor to be accurately calculated between a paper and a reviewer even if they do not share any keyword in common.

**Keywords:** taxonomy of keywords; semantic similarity; objects' description and classification; automatic assignment of reviewers to papers; conference management systems.

## 1. Introduction

One of the most important and challenging tasks in organizing scientific conferences is the assignment of reviewers to papers. It plays a crucial role in building a good conference image. For highly-ranked conferences, having a low acceptance ratio, it is very important that each paper is evaluated by the most competent, in its subject domain, reviewers. Even a small inaccuracy in the assignment may cause serious misjudgements that may dramatically decrease the conference image and authors' thrust in that event.

Generally the assignment could be done both manually and automatically. Manual assignment is applicable for small conferences having a small number of submitted papers and reviewers well known to the Programme Committee (PC) chairs. It requires the latter to familiarize themselves with all papers and reviewers' competences then assign the most suitable reviewers to each paper while maintaining a load balancing so that all reviewers evaluate

Yordan Kalmukov

roughly the same number of papers. Doing that for a large number of papers and reviewers is not just hard and time consuming, but due to the many constraints (accuracy, load balancing, conflict of interests and etc.) that should be taken into an account the manual assignment gets less and less accurate with increasing the number of papers and reviewers - a motive strong enough to force the developers of all modern commercially available conference management systems to offer an automatic assignment of reviewers to papers.

The non-intersecting sets of papers and reviewers can be represented by a complete weighted bipartite graph (figure 1), where P is the set of all submitted papers and R – the set of all registered reviewers. There is an edge from every paper to every reviewer and every edge should have a weight. In case of a zero weight, the corresponding edge may be omitted turning the graph to a non-complete one. The weight of the edge between paper $p_i$ and reviewer $r_j$ tells us how competent (suitable) is $r_j$ to review $p_i$. This measure of suitability is called a similarity factor. The weights are calculated or assigned in accordance with the chosen method of describing papers and reviewers' competences [7].

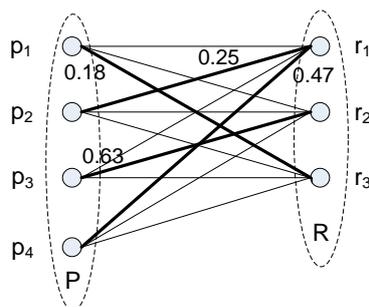

**Fig. 1.** The sets of papers (P) and reviewers (R) represented as a complete weighted bipartite graph. The edges in bold are the actual assignments suggested by an assignment algorithm. All edges have weights but just those of the assignments are shown for clearness.

Once the weights are calculated the automatic assignment could be easily implemented by using assignment algorithms known from the graph theory – for example the Hungarian algorithm of Kuhn and Munkres [8, 11] or heuristic algorithms like the one proposed in [5, 6]. Some advanced algorithms do really guarantee finding the best possible assignment for a given set of weighted edges, thus the weights' accuracy turns to be a key point in performing an accurate automatic assignment.

This article proposes a method of describing papers and reviewers' competences, based on hierarchically-structured set (taxonomy) of keywords. It significantly increases the weights' accuracy in comparison to other methods used in the existing conference management systems. The suggested similarity measure (equations 3 to 7) is derived from the well-known and widely-used Dice's coefficient, but it could be also applied over sets whose elements are semantically related to each other. It allows a non-zero similarity factor to be accurately calculated between a paper and a reviewer even if they do not share any keyword in common.

From now on the terms "conference topics" and "keywords" are used interchangeably.

## 2. Related Work

### 2.1. Conference and Journal Management

In respect to paper submission and review, conference management and journal management have pretty much in common. The key difference however is in the assignment of reviewers to papers. Conferences have a paper submission deadline. Once the deadline has been reached there are N number of submitted papers and M number of registered reviewers. The assignment is then handled as an assignment problem [8, 11, 5, 6] in bipartite graphs (as shown on figure 1). In contrast, journals usually do not have a submission deadline and papers are processed individually. The lack of clusters of papers makes the automatic assignment almost needless. Instead the journal management system should be able just to identify the most suitable reviewers (who are both competent and not overloaded) for a specified article without handling the assignment as a global optimization problem as the conference management system does. Despite that calculating similarity factors between a specific paper and all of the reviewers as accurate as possible is still very important for finding the most suitable reviewers to evaluate the paper.

One of the well known and probably the most used journal management system is Editorial Manager (EM) [27]. It is fully featured and highly configurable software that provides literally all editors need for flexible management. EM relies on hierarchical classification of terms to describe articles and reviewers' competences and to find the most competent reviewers. In contrast to the proposed solution however EM just takes into account the number of exactly matching classification nodes and omits the semantic relationships between them. Thus if the manuscript is described by a classification node and a reviewer has selected its child node (for example), then according to EM the article and the reviewer will have nothing in common as the number of matching nodes is 0. However as the node selected by the reviewer is a child node of the one describing the manuscript then the reviewer should be suitable to evaluate it – a statement based on the assumption that if the reviewer has an expertise in a specific field then he/she should have sufficient knowledge in the more general one as well. In this case EM suggestions seem not to be quite correct.

## 2.2. Describing Papers and Reviewers' Competences in Conference Management Systems

Most of the conference management systems (CMSs) are actually online, often cloud-based, conference management services. End users do not get any software but a completely hosted service. Thus analyzing existing solutions is mostly based on their official documentation and/or user experience. Some systems have very detailed documentation while others do not reveal any technical issues or algorithms at all.

A detailed review and comparative analysis of the existing methods of describing papers and reviewers' competences could be found in [7]. The phrase "method of describing papers and competences" includes general concepts, algorithms, data structures, mathematical formulas, proper organization of the user interface and etc. so that all these allow authors not just to submit a single file, but to outline what the paper is about and reviewers to state the areas of science they feel competent to review papers in [7].

Generally the methods of describing could be divided into two main groups:
- Explicit methods – require users to explicitly outline their papers and/or competences, i.e. to provide some descriptive metadata;
- Implicit methods – intelligent methods that automatically extract the required descriptive data from the papers and from reviewers' previous publications available on the Internet.

### 2.2.1 Implicit Methods of Describing Papers and Competences

The implicit methods of describing papers and competences use intelligent techniques to fetch the required metadata from the papers themselves or from the Internet. They usually perform a text analysis of papers' content and/or online digital libraries, indexes and others resources - DBLP [25], ACM Digital Library [20], CiteSeer [33], Google Scholar [28], Ceur WS [21] and etc.

Andreas Pesenhofer et al. [12] suggest that *the interest of a reviewer can be identified based on his/her previous publications available on the Internet*. The proposal suggests that the reviewer's name is used as a search query sent to CiteSeer and Google Scholar to obtain his/her publications. Then Euclidian distance (used as a measure of similarity) between the titles of every submitted paper and every reviewer's publication is calculated based on the full-text indexed feature vector [12].

Stefano Ferilli et al. [3] suggest a similar solution - *paper topics are extracted from its title and abstract, and expertise of the reviewers from the titles of their publications available on the Internet*. The proposed method assumes there is a predefined set of conference topics and *it tells which topics exactly apply to which papers / reviewers*. It is done by applying a Latent Semantic Indexing (LSI) [30] over the submitted papers' title and abstract and the titles of reviewers' publications fetched from DBLP [25].

Evaluated by the IEA / AIE 2005 conference organizers, the result showed a *79% accuracy on average. As to the reviewers the resulting accuracy was 65%* [3].

Marko Rodriguez and Johan Bollen [16] suggest that *a manuscript's subject domain can be represented by the authors of its references*. In comparison to the previous two methods this one does not exploit the titles and the abstracts, but the names of the authors who appear in the reference section. The approach uses a co-authorship network, initially built from the authors' names in the reference section of the paper; then their co-authors, the co-authors of the co-authors and etc. are fetched from DBLP, and a relative-rank particle-swarm algorithm is run for finding the most appropriate experts to review the paper.

Implicit methods are convenient and time saving but *they rely on external data sources on the Internet that are more or less inertial and contain sparse information.* New papers and articles are indexed with months delay and not by one and the same bibliographic index. That results in incomplete papers' and/or reviewers' profile. Rodriguez and Bollen report that for JCDL 2005 *89% of the PC members and only 83% of the articles with bid data were found in DBLP* [16]. Pesenhofer states, for ECDL 2005, *for 10 out of 87 PC members no publications have been retrieved from CiteSeer and Google Scholar* [12]. Obviously, for those not found within the bibliographic indexes, assignment was at random.

All commercially available conference management systems rely on explicit methods of describing papers and competences. Implicit methods however could be very useful for automatic detection of conflicts of interest and a couple of systems employ them for exactly that purpose.

### 2.2.2 Explicit Methods of Describing Papers and Competences

Explicit methods require authors and reviewers to provide additional metadata in order to describe their papers respectively competences. Existing CMSs use the following three different ways of providing such metadata:
- Bidding / rating papers
  (reviewers state their interest / preference to each paper individually);
- Choosing keywords from a predefined unordered set of conference topics;
- Combined - topics + bidding.

Bidding requires reviewers to browse the list of all papers and to indicate if they would like to review or not each one of them. Reviewers usually browse just titles and abstracts and state their willing to review by selecting an option from a drop down menu (figure 2).

Yordan Kalmukov

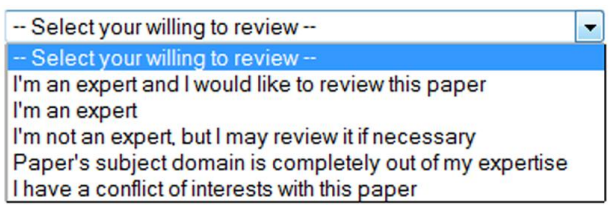

**Fig. 2.** A drop down menu allowing reviewers to indicate how willing they are of reviewing a specified paper

A similarity factor directly corresponds to each one of the options, where the highest factor is assigned to the highest level of willing to review. This method is considered to be the most accurate way of determining similarity factors as nobody knows better than the reviewer himself if he is competent to review the specified paper or not. It however does not really describe papers nor reviewers, but just the explicitly-stated relationship between them. If a paper has not been rated by enough reviewers then they will be assigned to it at random as there is no way the conference management system to determine the paper's subject domain or reviewers' competences.

Philippe Rigaux proposes a clever improvement of the bidding process which is trying to overcome the random assignments. The proposal is called an Iterative Rating Method (IRM) [15]. Its basic idea is to predict the missing ratings by iteratively applying a collaborative filtering algorithm to the ratings that has been explicitly provided by reviewers or the ratings that have been previously predicted (within previous iterations). The iterative rating method is fully implemented in the very popular The MyReview System [34].

Describing papers and reviewers' competences by a set of conference topics assumes that PC chairs set up a predefined list (unordered set) of keywords that best describe the conference coverage area. During paper submission authors are required to select those topics that best outline their papers. Reviewers do the same - during registration they are required to select the keywords corresponding to the areas of science they are competent in. Topics could be user-weighted or not. Non-weighted topics are usually selected from HTML checkboxes (figure 3). The user-weighted keywords allow users not only to state that a topic applies to their papers or area of expertise, but to indicate how much exactly it applies. They are usually selected from drop down menus.

Most of the conference management systems do not reveal how exactly topics are used in calculating similarity factors. Those that do, do not exploit them in the best possible way.

EasyChair [26] does not use any complex similarity measure but just counts the number of keywords in common. If a paper has more than one common topic with the PC member, it will be regarded as if he chooses "I want to review this paper". If a paper has exactly one common topic with the PC member, it will be regarded as "I can review it" [26]. This allows just 3 different levels of similarity – "entirely dissimilar" (0 topics in common); "slightly similar" (1 topic in common); "very similar" (2+ topics in common).

- ☐ Data modeling
- ☑ Data types & Data structures
- ☐ Data mining
- ☐ Algorithms & Problem solving
- ☐ Automata & state machines
- ☑ Artificial intelligence
- ☐ Computer graphics
- ☑ Computer vision

**Fig. 3.** Example of a list of non-weighted keywords

In contrast to bidding, describing objects by list of keywords provides an independent and stand-alone description of every single paper and reviewer that in most cases reduces the "random assignments" problem. There is no theoretical guarantee however that calculated similarity factors are 100% accurate as they are not directly assigned in accordance with subjective human judgments, but are more or less distorted by mathematics trying to represent the subjective human perception.

Combination of both topics and bidding aims to bring the advantages of the two previously discussed methods together and to reduce the influence of their disadvantages. The best sequence of actions is: 1) Based on the selected topics the conference management system suggests to each reviewer a small subset of let say 20 papers which should be the most interesting to him / her; 2) Reviewers rate the suggested papers; and 3) If a paper has been rated by a reviewer then his/her rating completely determines the similarity factor between them. Otherwise, if the paper has not been rated by the specified reviewer, the similarity factor is calculated based on the selected topics.

Although most of the CMSs support both topics selection and bidding not many of them follow this sequence of actions.

Based on topics matching the MyReview System [34] selects a subset of papers to be explicitly rated by each reviewer. Then missing bids are either predicted by the IRM or implicitly determined by topics matching (papers are described just by a single topic while competences by multiple topics). If the paper's topic matches one of the reviewer's topics the bid is set to "Interested". If there is no topic in common the bid is set to neutral ("Why not") [15, 34]. The latter seems not to be a very relevant decision as the "why not" bid still indicates minor competences in the paper's subject domain that could not be justified as there is no explicit bid and no common topic.

OpenConf [32] facilitates automatic assignment based on topics matching or bidding (pro version only) but not based on both simultaneously.

ConfTool [23], Confious [22], CyberChair [24, 17] and Microsoft CMT [31] state they are using both topics and bidding but there is no information how exactly they calculate the similarity factors between papers and reviewers.

EasyChair, as previously mentioned, do not use any complex similarity measure, but just counts the number of topics in common allowing only three levels of similarity.

CyberChair allows reviewers not only to state which topics they are competent in, but also to indicate how much exactly they are competent in. This "topics weighting" feature is supposed to increase the assignment accuracy.

GRAPE [2] system's fundamental assumption is to prefer topics matching approach over the reviewers' bidding one, based on the idea that they give assignments more reliability [2]. It uses reviewers' preferences just to tune the assignments.

### 2.2.3 A Brief Analysis of the Explicit Methods of Describing Used by the Existing Conference Management Systems

The existing explicit methods of describing papers and competences are most effective when they are used together. However if used separately or not in accordance with the already suggested sequence of actions, the accuracy of the assignment could not be guaranteed because of the following major reasons:
1. Bidding is the most reliable way of determining if a reviewer is competent to evaluate a specific paper. However in case of a large number of submitted papers, reviewers never bid on all of them. If a paper has not been explicitly rated by enough reviewers then *they are assigned to it at random*.
2. The non-structured (unordered) set of conference topics provides an independent stand-alone description of all papers and reviewers, however its size should be limited to a reasonable number of semantically non-overlapping elements. If there are overlapping topics then the author and the reviewer may select different ones although they have the same subject domain in mind. If that happens the similarity factor between them will be calculated as zero, i.e. *reviewers will be again assigned to the paper at random*.
3. The limit in the list's size results in less detailed keywords or in incomplete coverage of the conference topics. If keywords are too general they can not describe objects in details. If keywords are specific (detailed) enough their number may not be sufficient to cover the entire conference's area of science leading again to *assigning reviewers at random*.
4. Simple matching based similarity measures or the three-level similarity used in EasyChair do not allow similarity factors to be calculated precisely due to the *low number of distinguished levels*.

Solving random assignment requires solving the following:

1. Similarity factors should be calculated by complex similarity measures that allow factors to change smoothly within the range of [0, 1] on many distinguished levels.
2. A new method should be suggested that allows the conference to be described with many more *semantically-related* keywords. The larger number of keywords will provide more detailed and precise description of papers and competences, and will significantly decrease the probability of authors or reviewers not being able to find relevant keywords for them. Semantic relationship between separate keywords will allow accurate non-zero similarity factors to be calculated between a paper and a reviewer even if they do not share any keyword in common.
3. Existing similarity measures should be adapted or new ones should be proposed that not just count the number of matching (common) keywords, but calculates semantic similarity of the non-matching keywords as well.
4. Evaluation criteria should be proposed for assessing the accuracy of the suggested method of describing papers and reviewers' competences.

Solving task 1 is easy. As the list of conference topics is an unordered set the most reasonable way of calculating similarity factors is by using Jaccard's index (1) or Dice's similarity measure (2).

$$SF(p_i, r_j) = w(e_{p_i r_j}) = \frac{|KWp_i \cap KWr_j|}{|KWp_i \cup KWr_j|} \quad (1)$$

where:

$SF(p_i, r_j)$ - similarity factor between *i*-th paper and *j*-th reviewer
$KWp_i$ - set of keywords, describing the *i*-th paper
$KWr_j$ - set of keywords chosen by the *j*-th reviewer

$$SF(p_i, r_j) = w(e_{p_i r_j}) = \frac{2 \times |KWp_i \cap KWr_j|}{|KWp_i| + |KWr_j|} \quad (2)$$

Tasks 2 and 3 are solved by the suggested, in section 3, method of describing conferences, papers and reviewers' competences by taxonomy of keywords.

## 3. Describing Papers and Reviewers' Competences by Taxonomy of Keywords

The advantage of the proposed method comes from the hierarchical structure itself. It provides an important additional information – the semantic

Yordan Kalmukov

relationship between the separate topics (keywords) that allows similarity measures to take into account not only the number of keywords in common between a paper and a reviewer, but *in case of non-matching keywords to calculate how semantically close they are*. Thus if a paper and a reviewer do not share even a single topic a non-zero similarity factor could still be accurately calculated.

Before the paper submission and reviewer registration the conference chairs define taxonomy (figure 4) of topics that best describe the conference coverage area. The particular taxonomy on figure 5 is derived from the ACM classification scheme [19] and can be used to describe a software-related conference.

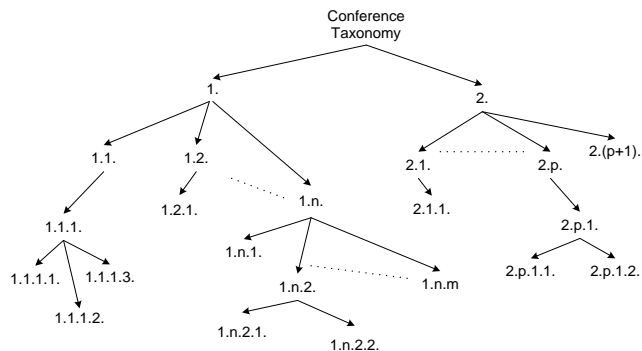

**Fig. 4.** General taxonomy structure

During paper submission authors are required to select all keywords that apply to their papers. During registration, reviewers select all topics that describe their areas of expertise.

Keywords describing the *i*-th paper and the *j*-th reviewer are stored into unordered sets noted as *KWp$_i$* and *KWr$_j$* respectively.

KWp$_i$ = {Relational databases, Content analysis, Web-based services, Architectures}

KWr$_j$ = {Relational databases, Distributed databases, Spatial DB & GIS, Information storage and retrieval, Content analysis, Data sharing, Software Engineering, Programming languages, C++}

Although keywords, chosen by authors and reviewers, are stored into unordered sets the semantic relationship between their elements is still available within the conference taxonomy. The most precise way of computing similarity factors is by using a measure of similarity between two sets of concepts in a common taxonomy.

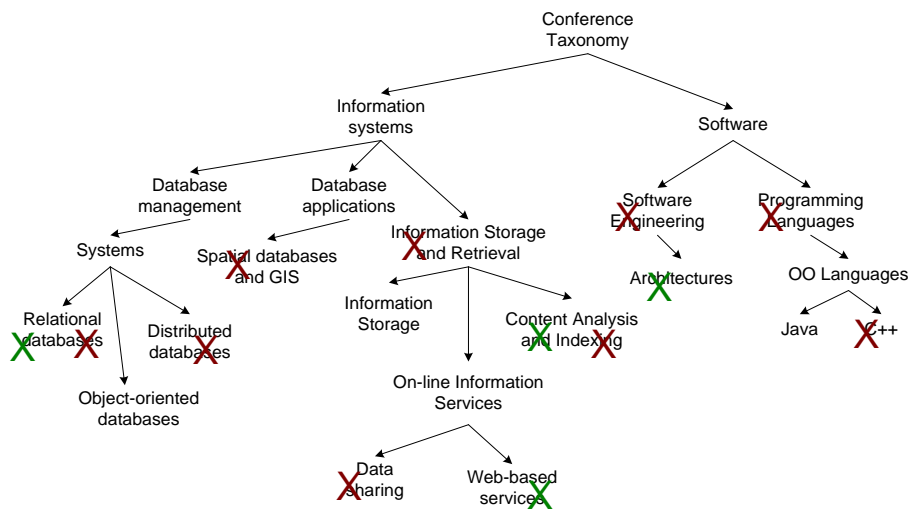

**Fig. 5.** An example for conference taxonomy with nodes chosen by the author of the i-th paper (coloured in green) and nodes chosen by the j-th reviewer (coloured in light brown)

There are many ways to compute similarity between two unordered sets (Dice, Jaccard and etc.), and many measures calculating the similarity between two individual concepts in a common taxonomy (Wu & Palmer [18], Lin [9], Resnik [14] and etc.). However there are just a few proposals for measuring similarity between sets of concepts in a common taxonomy and they are all asymmetric (Maedche & Staab [10], Haase et al [4]) or take into account all combinations between the elements of the two sets (Rada et al [13], Bouquet et al [1]) which is not applicable to the current subject domain.

The rest of this section proposes a way to compute similarity between two sets of concepts by combining similarity measures between sets with those measuring similarity between individual taxonomy concepts.

The original Dice's similarity measure is applicable to unordered sets only. Applying it to taxonomies will immediately convert the taxonomies to unordered sets as it measures commonality just by the number of exactly matching elements and ignores the semantic relationship between them. If a paper and a reviewer have no topics in common the similarity factor between them calculated by using (1) or (2) will be 0, although the reviewer could be competent to evaluate the paper. Consider for example a paper is described by "Object-oriented programming languages" and the reviewer has chosen "Java". In this case it is obvious the reviewer is competent to review the paper but as the topics do not exactly match the calculated similarity is zero.

To overcome that we can substitute the intersect in the numerator of (2) with the right-hand side expression in (3) that does not just count the number of exactly matching keywords, but calculates the *semantic commonality* between the keywords describing the specified paper and reviewer, so that if they do not share even a single common topic the numerator will not be 0 in case the reviewer is competent to evaluate the paper.

Yordan Kalmukov

$$|KWp_i \cap KWr_j| \Rightarrow \sum_{k_m \in KWp_i} \max_{k_n \in KWr_j}(Sim(k_m, k_n)) \qquad (3)$$

where:
$k_m$ – the *m*-th keyword chosen by the author of the *i*-th paper.
$k_n$ – the *n*-th keyword chosen by the *j*-th reviewer.
$KWp_i$ – set of keywords, describing the *i*-th paper.
$KWr_j$ – set of keywords chosen by the *j*-th reviewer.
$Sim(k_m, k_n)$ – semantic similarity between the *m*-th keyword chosen by the author of the *i*-th paper and *n*-th keyword chosen by the *j*-th reviewer.
$\max_{n \in KWr_j}(Sim(k_m, k_n))$ – semantic similarity between the *m*-th keyword chosen by the author and its closest neighbour (in the conference taxonomy) chosen by the reviewer.

$Sim(k_m, k_n)$ is calculated in a way that $Sim(k_m, k_n) \in [0,1]$. Thus the semantic similarity between the *m*-th keyword chosen by the author and its semantically closest neighbour chosen by the reviewer will lie within the same interval of [0,1].

If both the author of the *i*-th paper and the *j*-th reviewer have selected the same keyword, i.e. $m \equiv n$ then $Sim(k_m, k_n) = 1$ and $\max_{n \in KWr_j}(Sim(k_m, k_n)) = 1$

Thus

$$\sum_{k_m \in KWp_i} \max_{k_n \in KWr_j}(Sim(k_m, k_n)) = |KWp_i \cap KWr_j| \qquad (4)$$

in case $Sim(k_m, k_n) = 1$ for $\forall m \in KWp_i$ and $\forall n \in KWr_j$, i.e. if the author and the reviewer have chosen the same keywords.

*Furthermore if the taxonomy is converted to an unordered set (by ignoring the semantic relationship between its elements) then (4) is always true – unconditionally, because the similarity between $k_m$ and $k_n$ could be just 1 or 0. This proves substitution (3) is relevant and does not change the spirit of the Dice's measure (2).*

The intersection between two unordered sets is a commutative operation however the semantic commonality between two sets of concepts within a common taxonomy is not, thus:

$$\sum_{k_m \in KWp_i} \max_{k_n \in KWr_j}(Sim(k_m, k_n)) \neq \sum_{k_n \in KWr_j} \max_{k_m \in KWp_i}(Sim(k_n, k_m)) \qquad (5)$$

The reason (5) to be true is that the different number of elements within the two sets matters as each element has a non-zero similarity with an element from the other set.

In case of taxonomy, applying the suggested substitution (3) to the Dice's measure (2) results in (6). Note that the numerator is not twice the left-hand side of (5) but both sums are taken into account.

$$SF(p_i, r_j) = \frac{\sum_{k_m \in KWp_i} \max_{k_n \in KWr_j}(Sim(k_m, k_n)) + \sum_{k_n \in KWr_j} \max_{k_m \in KWp_i}(Sim(k_n, k_m))}{|KWp_i| + |KWr_j|} \quad (6)$$

where all notations are previously explained.

As proven earlier if the taxonomy is converted into an unordered set, by ignoring the semantic relationship between its elements, then (6) produces the very same result as the Dice's measure (2).

Equation (6), in contrast to the original Dice's measure, could be also applied between sets whose elements are semantically related to each other. Its asymmetric version is (7).

$$SF(p_i, r_j) = \frac{\sum_{k_m \in KWp_i} \max_{k_n \in KWr_j}(Sim(k_m, k_n))}{|KWp_i|} \quad (7)$$

where all notations are previously explained.

The symmetric similarity measure (6) takes into account:
a) how competent a reviewer is to evaluate a specified paper;
b) how focused the reviewer is on the paper's subject domain;
c) how much the paper's topics cover the reviewer's interests.

b) and c) show how worthy is to assign (spend) the reviewer for that paper exactly or it is better to keep him for other papers.

In contrast the asymmetric similarity measure (7) takes only into account how competent the reviewer is to evaluate the specified paper. So, which one to use – (6) or (7)? They both have their significance and which one applies better depends mainly on the assignment algorithm.

If reviewers are assigned to papers by a greedy algorithm then it is mandatory to adopt a symmetric measure. Greedy algorithms process papers in tern, assigning the most competent available reviewers to the current paper, but ignoring the possibility that another paper may appear later whose only suitable reviewers are already busy, i.e. overloaded and no more papers could be assigned to them. As a result the latter paper may remain without reviewers although there are competent ones to evaluate it but they have been previously and unreasonably used. In this case papers being processed

Yordan Kalmukov

first usually get the most competent, in their areas, reviewers however many reviewers may be assigned at random to the papers processed at last. The symmetric similarity measure (6) takes into account not just how competent the reviewer is, but also how worthy is to spend him for that paper exactly. In this way it significantly reduces the amount of random assignments to last processed papers.

If the assignment is handled as a global optimization problem then an asymmetric similarity measure could be used as well since the algorithm will find an optimal solution where there are competent reviewers for all papers.

To calculate the similarity factor between any paper and any reviewer by using (6) or (7), the software needs to know the semantic similarity between any two nodes in the taxonomy – the $m$-th keyword chosen by the author of the $i$-th paper and the $n$-th keyword chosen by the $j$-th reviewer, i.e. $Sim(k_m, k_n)$. There are two general ways of calculating semantic similarity between taxonomy nodes:

- by using the structural characteristics of the taxonomy – distance, depth, density and etc.
- based on the information content of the nodes.

A commonly accepted similarity measure based on structural characteristics is the one formulated by Zhibiao Wu and Martha Palmer [18] (8).

$$Sim_{Wu\ \&\ Palmer}(k_m, k_n) = \frac{2 \times N_0}{2 \times N_0 + N_1 + N_2}$$ (8)

where:

$N_0$ - the distance (in number of edges) between the root and the closest common ancestor $C_0$ of the two nodes ($C_m$) and ($C_n$) whose similarity is being calculated (Figure 6).

$N_1$ - the distance from $C_0$ to one of the nodes, say $C_m$. $C_m$ represents the $m$-th keyword describing the paper.

$N_2$ - the distance from $C_0$ to the other node – $C_n$. $C_n$ is the node representing the $n$-th keyword chosen by the reviewer.

As Wu and Palmer's measure is a symmetrical one it does not matter if $C_m$ is the node chosen by the author and $C_n$ the one chosen by the reviewer or the opposite. Closer look at (8) reveals it is in fact the Dice's coefficient applied over the definition of the term "path", i.e. a set of edges – twice the number of common edges divided by the number of all edges (including duplicates).

The calculated semantic similarity between any two nodes is only useful if it complies with the real human perception of similarity between these nodes. Dekang Lin [9] evaluated the Wu & Palmer's similarity measure by correlating the calculated similarity factors on 28 pairs of concepts in WordNet [35] with the assessments made by real humans during the Miller and Charles survey [14]. The experiment shows very good correlation (0.803) with the Miller and Charles results.

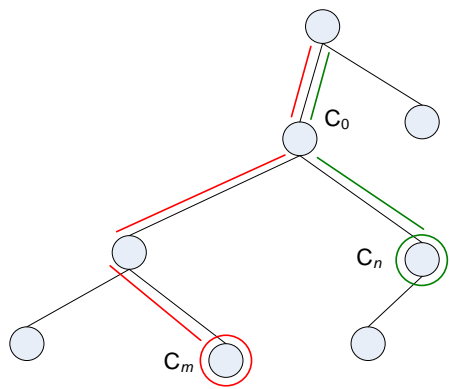

**Fig. 6.** Visual representation of Wu and Palmer's similarity measure.

Dekang Lin himself suggested another similarity measure based on the information content of the taxonomy nodes [9]. The taxonomy is augmented with a function *p:C -> [0, 1]*, such that for any node (concept) $c \in C$, *p(c)* is the probability of encountering the concept *c* or an instance of it. So if $c_1$ IS-A $c_2$, then $p(c_1) <= p(c_2)$. The probability of the root node (if any) is 1. Lower probability results in higher information content. Obviously and expectedly nodes deeper in the hierarchy are more informative than the ones on the top.

Lin's similarity measure [9] (9) looks similar to the one of Wu and Palmer, but takes into account the nodes' information content instead of distances.

$$Sim_{Lin}(k_m, k_n) = \frac{2 \times \log P(C_0)}{\log P(C_l) + \log P(C_m)} \qquad (9)$$

where:

$P(C_0)$ - the probability of encountering (in the conference taxonomy) the closest common ancestor $C_0$ (of the two nodes whose similarity is being calculated) or an instance of it;

$P(C_m)$ - the probability of encountering the node representing the *m*-th keyword chosen by the author or an instance of it;

$P(C_n)$ - the probability of encountering the node representing the *n*-th keyword chosen by the reviewer or an instance of it.

Lin's measure was part of the same experiment and evaluation that Lin performed on the Wu and Palmer's measure as well. It proved Lin's measure has a bit better correlation (0.834) with the real human assessments taken from the Miller and Charles data.

Yordan Kalmukov

In case of user-weighted keywords the semantic similarity calculated by using equations (8) or (9) could be additionally modified in the form of (10) to take the levels of competence and applicability into an account as well. These levels are usually explicitly provided by the users – authors and reviewers indicate how much exactly each one of the chosen keywords applies to their papers, respectively competences. The main idea behind this modification is to lower the semantic similarity in case the reviewer is less competent in respect to a specified keyword than the level it applies to the paper.

$$\left| \begin{array}{l} Sim_{weighted}(k_m, k_n) = Sim(k_m, k_n) \times C_{levels} \\ \text{If } w_{rj}(k_n) >= w_{pi}(k_m), C_{levels} = 1 \\ \text{else } C_{levels} = 1 - (w_{pi}(k_m) - w_{rj}(k_n)) \end{array} \right. \quad (10)$$

where:

$w_{rj}(k_n)$ - the expertise of reviewer $r_j$ on the topic $k_n$. $w_{rj}(k_n) \in [0, 1]$.

$w_{pi}(k_m)$ - the level that $k_m$ applies to paper $p_i$. $w_{pi}(k_m) \in [0, 1]$.

If both the author and the reviewer have selected one and the same keyword, i.e. $m \equiv n$, then:
- If the level of expertise (on that particular keyword) of the reviewer is higher than or equal to the level chosen by the author then the reviewer is considered to be highly competent to evaluate the specified paper *in respect to that particular keyword only*.
- If the level of expertise of the reviewer is less than the level chosen by the author, then the reviewer is less competent to evaluate the paper in respect to that keyword.

Logic is the same even if the author and the reviewer have selected different, but semantically close topics, i.e. $m \neq n$. If the modified semantic similarity $Sim_{weighted}(k_m, k_n)$ (10) is used in (6) or (7) instead of just $Sim(k_m, k_n)$ then it is possible for a semantically closest neighbour of the $m$-th keyword to be chosen not the nearest node selected by the $j$-th reviewer but another one for which the $j$-th reviewer has specified a higher level of expertise.

The system (10) takes into account the *relative level of competence*, i.e. whether the reviewer is more competent in $n$ than $m$ applies to the paper and if so the semantic similarity between $m$ and $n$ remains the same, otherwise it is lowered proportionally to the difference between the levels of $m$ and $n$. Alternatively the *absolute level of competence* of the reviewer could be used instead. In that case not the difference between the two levels matters but just the level the reviewer is competent in $n$. Thus (10) is transformed to (11).

$$Sim_{weighted}(k_m, k_n) = Sim(k_m, k_n) \times w_{rj}(k_n) \quad (11)$$

where all notations are previously explained.

## 4. Experimental Evaluation

A similarity factor should be calculated between any paper and any reviewer. The number of reviewers is usually proportional to the number of submitted papers so that reviewers' workload is kept in a reasonable range. Thus in respect to the number of submitted papers and registered reviewers the proposed method of describing the conference has a quadratic computational complexity, $O(n^2)$, where n – the number of submitted papers. The method of describing by an unordered set of conference topics has the same complexity as well, but significantly lower constant factor. The one suggested here reasonably runs slower due to the necessity of calculating semantic similarity between every keyword, $m$, describing the $i$-th paper and every one, $n$, chosen by the $j$-th reviewer. Fortunately the number of calculations could be significantly reduced by using optimization techniques like dynamic programming.

Evaluating accuracy of the methods of describing papers and reviewers' competences is a challenging task as they involve many subjective aspects like self-evaluation, self-classification, consistency of the taxonomy structure and etc.

The most reasonable way of assessing a method's accuracy is by comparing the calculated similarity factors to the similarity evaluations provided by real PC members over the same dataset. A reliable way of collecting reviewers' preferences is the method of bidding [7]. If the conference utilizes both the suggested method and bidding the similarity factors, calculated by the above equations could be compared to the bids provided by the reviewers. Five bid options are commonly available to reviewers to indicate their preferences and willing to review specific papers or not. These are:

1. I'm an expert in the paper's subject domain and I want to review it;
2. I'm an expert in the paper's subject domain;
3. I'm not an expert in the paper's subject domain, but I can review it;
4. The paper's subject domain is completely out of my expertise;
5. Conflict of interests.

Alternatively the calculated similarity factors could be compared to the reviewers' self-evaluation of expertise in respect to the papers they evaluate. The self evaluation of expertise is provided by each reviewer during the review submission, i.e. after reading the entire paper. It allows the conference management system to detect also errors of the following types "Selection of too general nodes", "Incomplete / inappropriate description of papers", "Ambiguous taxonomy structure" and others.

The proposed method of describing papers and reviewers' competences has been used by the international broad-area, multidisciplinary, ICT-related conference *CompSysTech* [29] in 2010 and 2011. During review submission every reviewer was required to specify his/her level of expertise in the paper's subject domain by selecting one of the following: **High**, **Medium** or **Low** level

Yordan Kalmukov

of expertise. Then the calculated similarity factors are compared to the reviewers' self-evaluations.

The method is considered to be accurate if most of the similarity factors are accurately calculated, i.e. comply with one of the following rules:

A similarity factor associated with **low** level of competences is correctly calculated if, for the paper it applies, it **is less than** all correctly calculated factors related to **medium** and **high** level of competences.

A similarity factor associated with **medium** level of competences is correctly calculated if it is **non-zero** and, for the paper it applies, it is **higher than** all correctly calculated similarities corresponding to **low** level of competences and **less than** all correctly calculated similarities corresponding to **high** level of competences.

A similarity factor associated with **high** level of competences is correctly calculated if it is **non-zero** and, for the paper it applies, it is **higher than** all correctly calculated similarities related to **low** and **medium** levels of competences.

In other words, in respect to a particular paper, the reviewer who stated to be an expert should have higher similarity factor than the one who is not an expert but still capable of reviewing the paper, and the latter should get higher similarity factor than the reviewer who is completely out of the paper's subject domain.

The relationship between the similarity factors on a local scale (per paper) is what really matters, not the specific values and not even the relationship on a global scale (for all papers simultaneously). This statement is true because the most competent reviewer to evaluate a particular paper is the one who has the highest similarity factor with the paper and no other papers can influence that similarity. Furthermore the actual value of the similarity factors depends on the number of chosen keywords while there is no rule stating that each paper or reviewer should be described by precisely the same number of topics.

The assignment algorithm used by the CompSysTech's conference management system is a heuristic but not greedy. It tries to find a global solution (although it does not guarantee finding the best possible one) and it does not assign reviewers to any paper until it finds suitable reviewers for all papers. That allows both symmetric and asymmetric similarity measures to be used. Since the reviewers' self-evaluation of expertise takes into account just the level of competence (which is an asymmetric in nature) it is expected that the relative amount of correctly calculated similarity factors will be similar for both the symmetric and the asymmetric measures. Results for CompSysTech'11 are summarized in table 1.

**Table 1.** Accuracy evaluation of the similarity factors of CompSysTech 2011

|  | **CST'11** (asymmetric measure) | **CST'11** (symmetric measure) |
|---|---|---|
| Number of assignments | 525 | 525 |
| Correctly calculated similarity factors | **83.24 %** | **83.05 %** |

Although the relationship between similarity factors on a global scale is not suitable for evaluation of the method's accuracy it could be very useful for identifying the reasons for having inaccurately calculated similarities. Figure 7 shows the distribution of the similarity factors of the three groups (Low, Medium and High) within the interval of [0, 1]. Line chart is used instead of bar chart as lines better illustrate trends. Furthermore in case of multiple data series on the same chart lines look much clearer than thick bars.

Despite the high extent of overlapping, similarity factors of the different groups tend to cluster where "High" have their maximum within the subinterval [0.8, 0.9), followed by "Medium" in [0.7, 0.8) and then "Low" in [0.6, 0.7).

Two "anomalies" are easily noticeable from the graphic:

- Relatively high amount of similarity factors associated with *High* level of competences within the lower subinterval of [0.3, 0.5];
- Relatively high mode value, 0.6-0.7, of the similarity factors associated with *Low* level of competences.

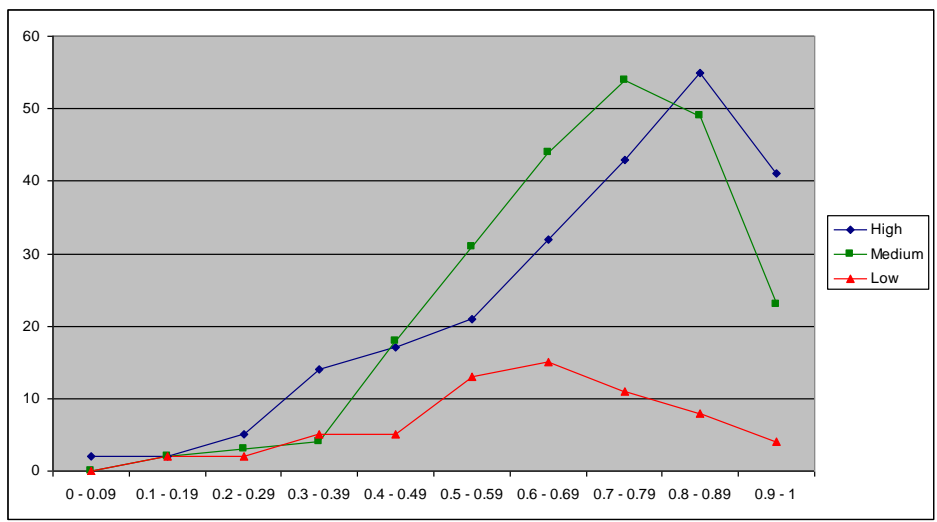

**Fig. 7.** Distribution of similarity factors (calculated by an asymmetric measure) of the three groups (Low, Medium and High) within the interval of [0, 1] for CompSysTech'11.

A detailed review of the inaccurately calculated similarity factors reveals the following major reasons for having inaccuracies:

Yordan Kalmukov

- ***Incomplete description of competences*** - reviewers choose keywords just within a narrow area of science, but during the budding phase they bid as experts on papers outside that area.
- ***Incomplete description of papers*** – multi-disciplinary papers described by keywords from just one area. For example: e-learning on hardware subjects, but described by e-learning keywords only.
- ***Selection of two general, low-informative nodes***. For example: Software or Computer Applications. The lack of details does not allow papers and competences to be described precisely.
- ***Multi-disciplinary papers with higher contribution in a non-primary area*** (non-computing area in our context) which the reviewer could not evaluate without having a sufficient knowledge in that area. For example, with computing as a primary area, papers related to yoghurt processing, internal combustion engines, medicine, finance and etc.
- ***Ambiguous taxonomy structure.*** For example duplicate keywords located within different nodes in the tree.

Incorrectly calculated similarity factors not due to the specified reasons are marked as "Unclassified". Their nature is more or less random and could be a result of other, unforeseen, subjective factors or inappropriate taxonomy structure. It is preferable taxonomy to be narrow and deep rather than broad and shallow. Figures 8 and 9 show the contribution of the individual reasons to the presence of *High* competences associated with low similarity factors (fig. 8) and *Low* competences associated with high similarity factors (fig. 9).

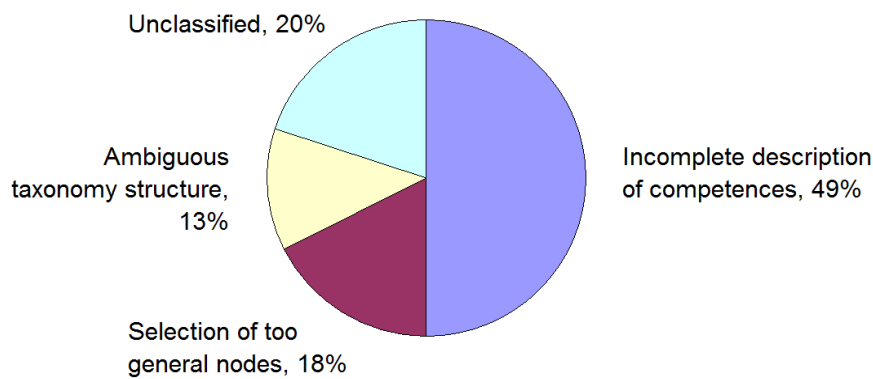

**Fig. 8.** Contribution of the individual reasons to the presence of low similarity factors (< 0.5) associated with *High* level of competences.

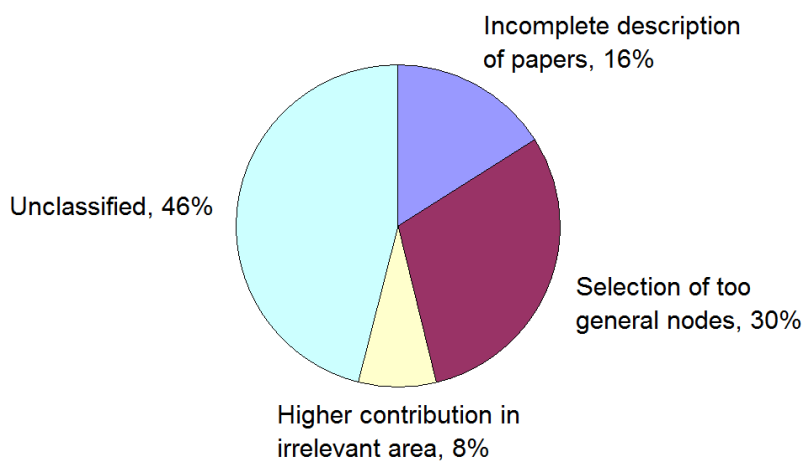

**Fig. 9.** Contribution of the individual reasons to the presence of high similarity factors (> 0.5) associated with Low level of competences.

Since the major causes of inaccuracy are identified they may be partially or completely avoided and/or compensated. Refer to section "Conclusions and Future Improvements" for more information.

### *Comparing the suggested method to the one describing papers and competences by an unordered set (often referred as list) of keywords*

Comparing the accuracy of two or more explicit methods is a challenging task due to the subjective aspects of self-classification. *The fairest comparison* requires that different methods are implemented in a single conference management system so *they are used in parallel, in one and the same event, and by the same authors and PC members – a situation that is more or less unrealistic.*

Fortunately alternative ways of indirect comparison are also possible. For example CompSysTech 2009 was described by an unordered set of keywords but the assignment was done by the same assignment algorithm, configured in the very same way and the Programme Committee in 2009 was the same as in 2010. In 2011 PC was larger but contained all members from 2009 as well. Then besides the papers the only other difference is the method of describing papers and competences. Everything else (the assignment algorithm and the PC members) is the same. That allows us to indirectly compare the two methods.

Table 2 shows the relative amount of correctly calculated similarity factors (according to the accuracy rules stated earlier) for CompSysTech 2009. Expectedly the asymmetric measure shows a little bit higher result due to the asymmetric natures of the reviewers' self-evaluation of expertise. The asymmetric measure is calculated as the number of common keywords

Yordan Kalmukov

divided by the number of keywords describing the paper, while the symmetric measure is the Jaccard's index (1).

**Table 2.** Accuracy evaluation of the similarity factors of CompSysTech 2009

|  | **CST'09** (asymmetric measure) | **CST'09** (symmetric measure) |
|---|---|---|
| Number of assignments | 356 | 356 |
| Number of random assignments (zero calculated similarities) | 64 | 64 |
| Number of zero-calculated similarity factors associated with High and Medium level of competences | 57 | 57 |
| Correctly calculated similarity factors | **67.70 %** | **67.13 %** |

As discussed in section 2 the method of describing papers and competences by unordered set of keywords is characterized by a large number of random assignments, i.e. zero-calculated similarity factors. As seen on the table most of them are actually related to medium and high level of competences. Although the zero-calculated similarities, reviewers are assigned to the corresponding papers just because they (reviewers) have directly chosen the papers they would like to evaluate. If reviewers were unable to do that, papers should have been assigned to them at random, significantly decreasing the overall number of assignments associated with high and medium level of expertise.

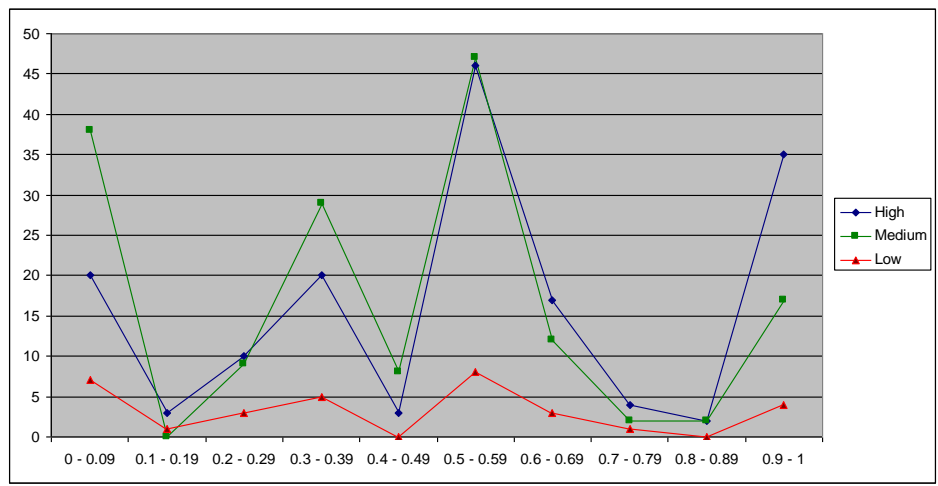

**Fig. 10.** Distribution of similarity factors (calculated by an asymmetric measure) of the three groups (Low, Medium and High) for CompSysTech'09.

The global scale distribution of similarity factors, calculated by the asymmetric similarity measure, for CompSysTech 2009 is shown on figure 10.

In contrast to the bell-shaped like distribution produced by the suggested method, the asymmetric similarity measure of the one using unordered set of keywords provides reduced set of values together with more chaotic distribution. As seen on the figure all the three groups (Low, Medium and High) have their maximums in the same subinterval of [0.5, 0.6). However after 0.6 similarity factors associated with *High* level of competences are more than the ones corresponding to *Medium*, i.e. the unordered set of keywords could be also used to describe papers and competences although it provides less precision and accuracy, in comparison to the taxonomy, especially for broad area and/or multi-disciplinary conferences.

One may say the accuracy levels in 2011 and 2009 are just lucky - accidentally achieved. Table 3 contradicts that statement by showing accuracy levels for all years from 2008 to 2011. For both years when the conference management system used taxonomy for describing papers and competences the accuracy levels are higher than the two years when the system relied on an unordered set of keywords. In 2008 papers were evaluated by just two reviewers that improperly make the overall accuracy higher (the fewer number of similarity factors per paper decreases the possibility that they are wrong since one of the similarities is taken as a reference value, i.e. considered to be correct) but it is still under 70%.

**Table 3.** Accuracy evaluation of the similarity factors of CompSysTech 2008 to 2011 (asymmetric similarity measures)

|  | **CST'11** | **CST'10** | **CST'09** | **CST'08** |
|---|---|---|---|---|
| Method of description | taxonomy | taxonomy | unordered set | unordered set |
| Reviewers per paper | 3 | 3 | 3 | 2 |
| Number of assignments | 525 | 385 | 356 | 196 |
| Number of random assignments (zero similarities) | 0 | 0 | 64 | 35 |
| Correctly calculated similarity factors | **83.24 %** | **80.26 %** | **67.70 %** | **68.88 %** |

## 5.   Conclusions and Future Improvements

The suggested method relies on taxonomy of keywords (conference topics) used as descriptive metadata. The implied hierarchical structure allows similarity measures to calculate similarity between a paper and a reviewer even if they do not share any common keyword at all. That avoids random assignments and increases the weights' accuracy in general, resulting in more precise automatic assignment.

The method however does not fully eliminate the inaccurately calculated similarity factors. Since most of the reasons for having inaccuracies are identified during the analysis they could be completely or partially avoided and/or compensated as follows:

1. To avoid selection of too general, low informative nodes (ex. Computer Applications; Software and etc.) it is enough to technically disallow users to select these nodes (or all nodes within the first one or two levels after the root).

2. Partial description of competences could be significantly compensated by using *collaborative filtering* techniques. If a reviewer bids as an expert on papers described by keywords he/she has not chosen, then they could be automatically added to his/her set of keywords. Additional constraints may be applied here to avoid improper propagation of keywords due to incorrect selection by authors - for example keywords could be added to the reviewer's set only if: they are chosen by other reviewers who bid as experts on the same paper (proving the author's selection of keywords is correct); or if the keywords also describe other papers the reviewer has bid on as an expert (proving the reviewer has competences within the specified area but missed to mark the relevant keywords).

3. In some cases reviewers tend to generalize their competences and instead of selecting all nodes within a specific narrow area of science reviewers select just their (the nodes) common ancestor stating high level of competence on it. For example: If the reviewer feels highly competent on all aspects of information systems then he/she is more likely to choose the general node "Information Systems" rather than its sub-nodes. Achieving higher accuracy however requires users to select nodes deeper in the hierarchy. The following rule helps increasing the accuracy while providing reviewers' comfort by allowing them to generalize their competences by selecting just a single node.

*If*
    1. A reviewer has stated *high level of competence* for a node $n_i$
and 2. $n_i$ *is between* […] and […] level in the hierarchy
and 3. $n_i$ *has children*
and 4. the reviewer **has not** selected any of $n_i$'s children.
*then:*
Add $n_i$'s direct successors to the set of keywords chosen by the reviewer (assuming the reviewer is generalizing his knowledge and instead of selecting all children he has selected just their common ancestor for convenience)

Although the proposed method achieves higher accuracy in comparison to other methods used in the existing conference management systems its accuracy could be further increased by implementing the three proposals described above. Additionally the method should be verified with specialized narrow-area conferences where its accuracy is expected to be even higher.

A comparable solution based on paper-to-paper similarity could be used to group papers in working sessions.

The suggested, in this article, method and all accompanying formulas have been exclusively designed to precisely calculate similarity between papers and reviewers however they may be used to describe any other objects and subsequently to compute similarities between them – for example men and women on dating sites; or job vacancies and candidates and etc.

**Acknowledgement.** This paper is supported by project: Creative Development Support of Doctoral Students, Post-Doctoral and Young Researches in the Field of Computer Science, BG 051PO001-3.3.04/13, European Social Fund 2007–2013, Operational Programme "Human Resources Development"

Yordan Kalmukov

**Yordan Kalmukov** received his BSc and MSc degrees in Computer Technologies at the University of Ruse in 2005 and 2006 respectively. Currently he is a PhD student at the Department of Computer Systems and Technologies, University of Ruse and teaches BSc students in web programming. His research interests include information retrieval, automated document management and classification systems, web-based information systems and databases.